# Enhanced Thermoelectric Performance of Nanostructured Nickel-doped Ag$_2$Te

Vikash Sharma[1,†,*], Divya Verma[2,3], Ranu Bhatt[4], Pankaj K. Patro[5], Gunadhor Singh Okram[1,*]

[1]UGC-DAE Consortium for Scientific Research, Indore, Madhya Pradesh-452001, India

[2]Government College Alote, District Ratlam, Madhya Pradesh-457114, India

[3]Department of chemistry, Vikram University, Ujjain, Madhya Pradesh-456010, India

[4]Technical Physics Division, Bhabha Atomic Research Centre, Mumbai-400085, India

[5]Powder Metallurgy Division, Bhabha Atomic Research Centre, Vashi Complex, Mumbai-400703, India

[*]vikash.sharma@tifr.res.in, okram@csr.res.in

**ABSTRACT** We report on the thermoelectric properties of nickel-doped Ag$_{2-x}$Ni$_x$Te (x = 0, 0.015, 0.025 & 0.055, 0.115, 0.155) nanostructures in the temperature ($T$) range of 5 K to 575 K. The electrical resistivity ($\rho$) of Ag$_2$Te nanostructure shows metallic behaviour in 5 K to 300 K initially that evolves into two metal to insulator transitions (MITs) at low and mid-temperature regimes with increasing x due to Mott-variable range hopping (VRH) and Arrhenius transports, respectively. Their Seebeck coefficient varies nearly in a linear fashion in this temperature range, showing metallic or doped-degenerate semiconducting behaviour. Notably, this behaviour of Seebeck coefficient ($S \propto T$) is in contrast to Mott-VRH conduction ($S \propto T^{1/2}$) as observed in $\rho$. The steady increase in $\rho$ and $S$ with the sharp decrease in thermal conductivity between 410 K to 425 K associated with the structural phase transition accomplishes a maximum thermoelectric figure of merit (*ZT*) of 0.86±0.1 near 480 K in x = 0.155. This is ~ 83 % more compared to that of bulk $Ag_2Te$, and shows a significant improvement over the best value reported for $Ag_2Te$ nanostructures thus far. This study, therefore, shows that simultaneous nanocomposite formation, doping and nanostructuring could be an effective strategy for tuning the electron and phonon transports to improve the thermoelectric properties of a material.

**INTRODUCTION**

Thermoelectric (TE) materials can be used to generate electricity directly from temperature gradient in a reliable way that has the potential for recovery of waste heat as electricity as well as to overcome the current global energy crisis and future energy problems[1,2,3,4]. The conversion efficiency of a TE material or device can be examined using dimensionless figure of merit, $ZT = \frac{S^2\sigma}{\kappa}T$, where $\sigma$ is electrical conductivity, and $\kappa$ is the total thermal conductivity contributions from its electronic ($\kappa_e$) and lattice ($\kappa_l$) parts and *T* is the absolute temperature[5]. The possible enhancement of *ZT* through the large electrical conductivity and Seebeck coefficient with the minimum thermal conductivity is inherently in conflict in the usual semiconductors or metals. This is possible fortunately in nanostructures (NSs) by independently tuning the power factor (PF) $S^2\sigma$ and $\kappa_l$ as evident from, say, the size and morphology-controlled complex materials[6], nanocomposites[7,8], hybrid materials with inorganic/organic interfaces[9,10], introduction of nanoparticles (NPs) in host matrix[11], alloying and doping[12].

Silver telluride ($Ag_2Te$) is an interesting and attractive nonmagnetic topological insulator at ambient conditions[13,14] with many intriguing properties such as the structural phase transition from the low-temperature monoclinic phase β-$Ag_2Te$ to high temperature face-centred cubic (fcc) phase α-$Ag_2Te$ near 417 K[15,16,] pressure-induced charge density wave (CDW) phase[17], structural[17,18] and electronic topological[19] phase transitions. While α-$Ag_2Te$ phase is a superionic conductor, its β-phase is a narrow-band gap (~ 0.05 eV) semiconductor with high carrier mobility and low $\kappa_l$ due to the Ag-atoms-induced disordered structure in the $Ag_2Te$ lattice[20] making it a suitable TE material[16,15,21]. Although its low effective mass (~ $10^{-2}$ of free electron mass)[16] favours the small $S$, high mobility makes both $\sigma$ and $\kappa_e$ large. As a result, $ZT$ of ~ 0.27 at 370 K[22] and ~ 0.29 at 550 K[23] for bulk $Ag_2Te$ along with several attempts to improve upon its TE properties via nanostructuring, alloying and or doping[20,23,24,25,26,27,28,29] have been reported so far. Cadavid *et al.*[29] showed a peak $ZT$ of ~ 0.66 at 450 K in surface-functionalized $Ag_2Te$ NPs. Zhou *et al.*[23] reported $ZT$ of ~ 0.62 and ~ 0.47 at 550 K in sulphur-doped 15 nm NP sample and sulphur-doped bulk $Ag_2Te$, respectively. Yang *et al.* observed a peak $ZT$ of ~ 0.55 at 400 K in $Ag_2Te$ nanowires[20]. A highest $ZT$ of ~ 0.68 at 573 K in $Ag_2Te$/Ag nanocomposite[30] and $ZT$ of ~ 0.067 and ~ 0.32 at 300 K in uncapped and 1,2-ethanedithiol-capped $Ag_2Te$ nanocrystals, respectively have also been reported[25]. The maximum PF of ~1370 $\mu Wm^{-1}K^{-2}$ at 425 K in $Ag_{1.99}TeSe_{0.01}$/Ag nanocomposites[24], ~ 315 $\mu Wm^{-1}K^{-2}$ at 410 K in $Ag_2Te$ nanowire films[27], and ~ 3.94 $\mu Wm^{-1}K^{-2}$ at 300 K in $Ag_2Te$/Ag nanofibers[28] have been reported. Although these reports focus only on TE properties for the temperature over 300 K, whereas low temperature TE properties of this compound are not explored so far.

Here, we demonstrate the TE properties of $Ag_{2-x}Ni_xTe$ (x = 0, 0.015, 0.025, 0.055, 0.115, 0.155) NSs in the temperature range of 5 K to 575 K. The metallic behaviour of undoped nanostructure over T range of 5 K to 300 K, like bulk $Ag_2Te$[17], turns to two clear metal to insulator transitions (MITs) for x = 0.115 and 0.155. Their structural phase transition from β-$Ag_2Te$ to α-$Ag_2Te$ near 420 K is clearly evident in all $\rho$, $S$ and $\kappa$. Even though the peak PF of ~ 1254 $\mu Wm^{-1}K^{-2}$ at 580 K for x = 0 is larger than that of ~ 1175 $\mu Wm^{-1}K^{-2}$ at 420 K for x =

0.155, the latter shows the maximum *ZT* of ~ 0.86 compared to ~ 0.43 at 580 K of the former; the *ZT* of ~ 0.86 is associated with an ultralow value of $\kappa_l$ ~ 0.15 Wm$^{-1}$K$^{-1}$ at 482 K. This *ZT* value is significantly larger compared to achieved value in undoped bulk and nanostructured Ag$_2$Te[20,23,25,29,30]. This is attributed to the point defects, structural distortion, and anharmonicity prevailed as x increases, leading to the substantial optimization of the electron and phonon transports.

**EXPERIMENTAL SECTION**

Silver nitrate AgNO$_3$ ($\geq$ 99.0%), nickel chloride NiCl$_2$ (98%), potassium tellurite monohydrate K$_2$O$_3$Te.H$_2$O ($\geq$ 90%), and diethylene glycol (DEG, 99%) were used as received without further purifications. Typically, 2:1.1 of AgNO$_3$ and K$_2$O$_3$Te.H$_2$O was mixed in 30 ml of DEG in a three neck round bottom flask. This mixture was heated with rate of ~ 10 $^o$C/min up to ~235 $^o$C. The reaction temperature of ~235 $^o$C was maintained for 3 hrs, and product was cooled down to room temperature. Ethanol was added in the reaction product and centrifuged it at 12,000 rpm for 10 minutes. The resultant supernatant was discarded after retaining the sample. This washing step was repeated for three times and product was then vacuum-dried. The obtained powder sample was used for further characterizations. This sample gives x = 0. To prepare the Ni-doped nanostructured Ag$_2$Te, desired amount of NiCl$_2$ was replaced for AgNO$_3$ and followed similar preparation procedure for the nanostructured Ag$_{2-x}$Ni$_x$Te (x = 0.015, 0.025, 0.055, 0.115, 0.155) samples; values of x shall denote samples in the text. Further, doping of Ni with x = 0.205 led to the formation of the Ag$_2$Te/Ni$_{1.4}$Te nanocomposite.

Bruker D8 Advance X-ray diffractometer, TECHNAI-20-G$^2$ (200 KV), FEI Nova nanosem450, X-ray photoelectron spectroscope (SPECS, Germany) with Al K$\alpha$ radiation and SEM equipped with EDX JEOL JSM 5600 were used for x-ray diffraction (XRD), transmission electron microscopy (TEM), field emission scanning electron microscopy (FESEM), X-ray photoelectron spectroscopy (XPS) and energy dispersive analysis of x-ray (EDX), respectively.

A commercial Labram-HR800 micro-Raman spectrometer and vibrating sample magnetometer (VSM) of Quantum Design were used for Raman and magnetization measurements, respectively. Four probe and differential direct-current methods were used to measure the resistance and Seebeck coefficient over the $T$ range of 5–320 K in a home-made setup[31]. We have also measured TE properties in the $T$ range 300 K to 580 K. Linseis make LSR-3 measured Seebeck coefficient and electrical resistivity (four-probe method) simultaneously on the cold-pressed sintered (at 500 °C) rectangular pellets with dimension 5 mm (l) × 3 mm (b) × 2 mm (t). Dynamic laser flash technique using Linseis make LFA-1000 measured the thermal diffusivity ($D$) of the disc-shaped samples with a diameter of ~ 10 mm and thickness of ~ 2 mm to determine their thermal conductivities using the formula $\kappa = DC_p d$, where $C_p$ is specific heat and $d$ is mass density of the sample; the cold-compressed pellets were sealed in a quartz tube under a vacuum of ~ $10^{-6}$ torr, and annealed it at 500 °C for 24 hrs in a conventional furnace. Typical uncertainty in the measurements of $S$, $\rho$ and $\kappa$ were less than 3, 4 and 10%, respectively. The mass densities of these samples were greater than 82 % of the bulk value (8.2 g/cm$^3$) as listed in table S1.

**RESULTS AND DISCUSSION**

The XRD pattern of x = 0 confirms its monoclinic crystal structure with space group P21/c of β–Ag$_2$Te (figure 1a). The samples Ag$_{2-x}$Ni$_x$Te with x = 0.015, 0.025, 0.055, 0.115 and 0.155 retain the same β–Ag$_2$Te phase with a small fraction (< 3 %) of elemental Ag metal (figure S1b-f) that manifests the successful substitution of Ni up to about 15.5 % in the host Ag$_2$Te lattice (figure 1a). Therefore, these Ni-doped Ag$_{2-x}$Ni$_x$Te samples are effectively nanocomposites of Ag (<3%) and Ag$_2$Te (>97%). Their Rietveld refinements are shown in figure S1. XRD of x = 0.205 (figure S2) shows additional peaks corresponding to Ni$_{1.4}$Te (P4/nmm, S.G. # 129) along with main phase of Ag$_2$Te. This suggests that further doping of Ni cannot be possible beyond x = 0.155, and forms the Ag$_2$Te/Ni$_{1.4}$Te nanocomposite. The crystallite size calculated from Scherrer formula increases with increase in x from 0, 0.015, 0.025, and 0.055, except a decrease in x = 0.115 and 0.155 (table S1).

The β–$Ag_2Te$ has highly distorted antifluorite structure consisting of a triple layered Te (Ag)– Ag –Te (Ag) stacking structure wherein Te atoms occupy a distorted fcc lattice with Ag atoms inserted in the interstitials (figure 1a, inset)[18]. The substituted Ni atoms can occupy the Ag1 and/ or Ag2 sites that can lead to change in physical properties of the host $Ag_2Te$. The obtained lattice parameters a, b & c from Rietveld refinements of x = 0 - 0.155 are shown in figure 1b and table S1. While lattice parameter a or c decreases exponentially (figure 1b) or linearly

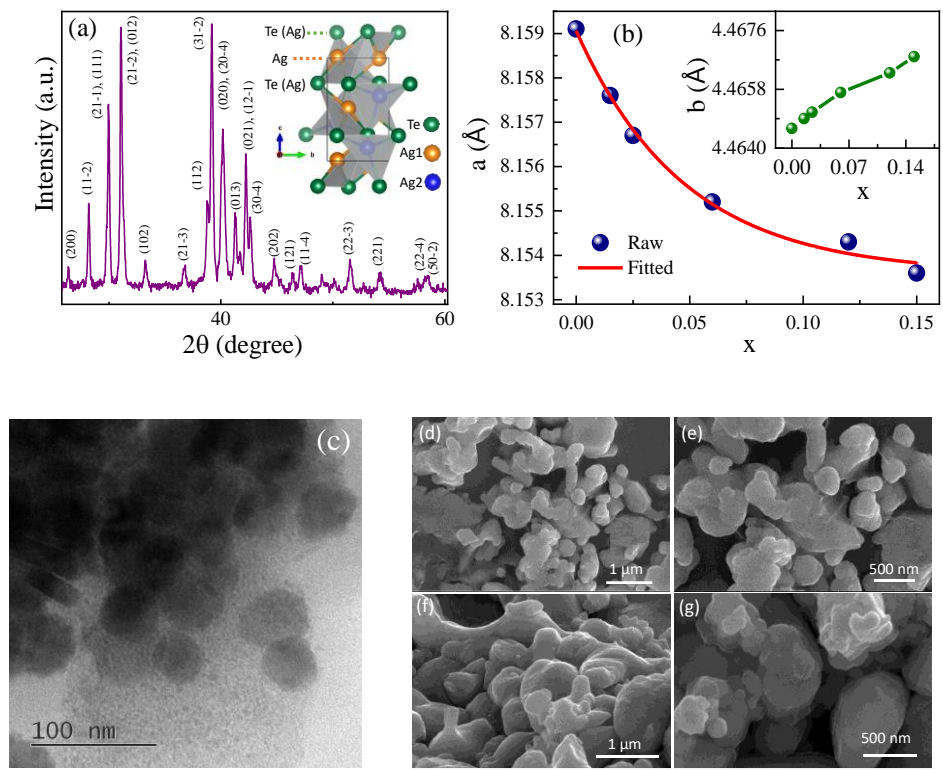

**Figure 1** (a) X-ray diffraction pattern of x = 0; inset shows its crystal structure of $Ag_2Te$, (b) variation in lattice parameter a with x; inset shows for lattice parameter b. (c) Transmission electron microscopy images of x = 0, scale bar is 100 nm. Field emission scanning electron microscopy images of $Ag_{2-x}Ni_xTe$ for (d, e) x = 0 and (f, g) x = 0.155.

(table S1), respectively, b increases marginally with x (figure 1b, inset). They show the signature of substitution of smaller radius Ni atoms for bigger Ag atoms. This matches with softening in the Raman modes in x = 0.155 compared to x = 0 (figure S3).

Figure 1c shows a representative TEM image of x = 0 in which particle size ranges from 30 nm to 55 nm with average particle size of ~ 42 nm; see figure S3 also. Morphology of x = 0 and x = 0.155 (figure 1d-g) are determined using FESEM. The random shaped clusters of size ranging from 200 nm to 1 µm can be seen in x = 0.00 (figure 1d, e). Clusters of size ranging from 400 nm to 1 µm are found in x = 0.155 (figure 1f, g). The clusters are relatively agglomerated in x = 0.155 compared to x = 0 with somewhat similar morphology, which is attributed to the change in nucleation and growth in the presence of additional precursor $NiCl_2$ even in the similar reaction conditions. The XPS reveals the presence of Ag and Te in $Ag^{+1}$ and $Te^{-2}$ valence states, respectively, without any impurity phase on the surface of x = 0 (figure S5). The EDX spectroscopy was employed to determine the stoichiometry of the synthesized samples. The atomic ratio of Ag and Te as 66.81 % and 33.19 %, respectively was found in x = 0 without any other impurity element (figure S6a). The EDX spectra of x = 0.015 - 0.155 are also shown in figure S6b-f, and their atomic percentage is listed in table S2; the details are discussed in SI. The EDX measurements were performed at different sites of x = 0.055 (figure S7a-c). The difference in atomic percentage of Ag, Ni and Te at different sites does not exceed 1 % that indicates reasonably good homogeneity of this sample; the same is expected in other samples as well.

Figure 2a-c shows the electrical resistivity $\rho$ measured on the cold-pressed sintered pellets of $Ag_{2-x}Ni_xTe$, x = 0 - 0.155 having mass density greater than 82 % of the bulk value (8.2 g/cm$^3$) (table S1) over the *T* range of 5 K to 300 K. The $\rho$ of x = 0 shows a slight increase below ~ 15 K but rises above it with increase in *T* like a doped narrow-bandgap semiconductor due to perhaps slowly decreasing carrier concentration (*n*) and phonon scattering[17]. Similar feature of $\rho$ for x = 0.015 with change in slope near 150 K and 240 K, and magnitude is seen (figure 2a). These features i.e. upturn at low $T$, a broad hump in mid-$T$ range and a broad dip at higher temperature become more pronounced and shift towards higher temperature with increase in x, evolving the overall behaviour from x = 0.015 to 0.055 quite dramatically (figure 2a, b). The electrical transport behaviour is significantly transformed in x= 0.115 and 0.155 such that the $\rho$ of x= 0.115 shows metallic behaviour between 300-230 K, semiconducting behaviour between 230-145 K, and again metallic behaviour between 145-45 K and finally semiconducting behaviour below ~ 45 K (figure 2c). These features are more pronounced in x= 0.155 with semiconducting behaviour below ~ 50 K and between 130 K - 200 K, and metallic behaviour between 50 K - 130 K and 210 K - 300 K (figure 2c). To assess the correlation of the MITs with any magnetic transitions, if any, and understand the short-range

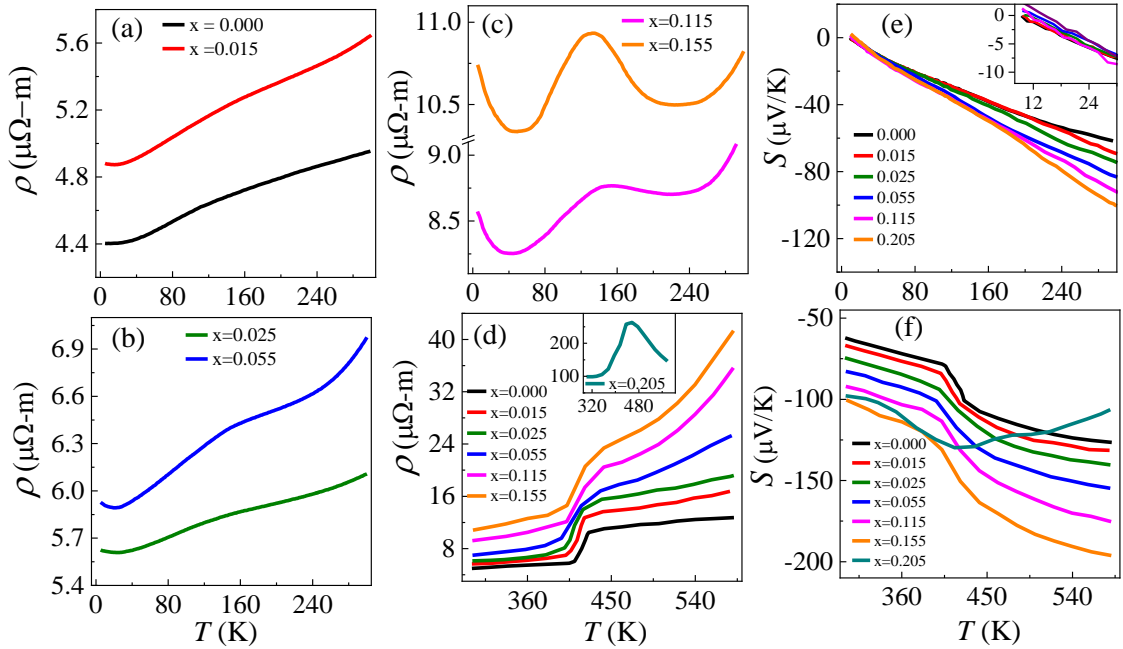

**Figure 2** Electrical resistivity of $Ag_{2-x}Ni_xTe$ for (a) x = 0 and 0.015, (b) x = 0.025 and 0.055, (c) x = 0.115 and 0.155 between 5 K to 300 K. (d) Resistivity of x = 0 - 0.155; inset shows $\rho$ for x = 0.205, (e) Seebeck coefficient for x = 0 - 0.155 between 5 K to 300 K and (f) Seebeck coefficient for x = 0 - 0.205 between 300 to 575 K.

correlation between Ni moments, magnetic measurements were carried out on x = 0.015 and 0.025 as representative samples. They provide no clear evidence of any magnetic transitions related to the MITs but shows prevalence of the possible superparamagnetism (figure S8); see details in SI. They indicate the rather complex electrical transport of MITs in different $T$ regimes (figure 2a-c), and are tentatively understood to be due to the interactions of the silver metal (< 3 %) with β–$Ag_2Te$ phase (figure S1b-f) in the presence of extensive grain boundaries (GBs) and nanostructures (figure 1c - g) since the magnitude of resistivity is rather low showing perhaps significant influence of silver at the GBs.

Such MITs may be due to the change in structure, CDW phase transition, and electron-electron correlation; CDW transitions are usually associated with opening of a gap near Fermi level. This is in line with earlier report on $Ag_2Te$[17]. There are changes in crystal structure across these CDW transitions, corresponding to perhaps to the lattice contraction as the chemical pressure. The low temperature transition might be due to electron-electron correlation. However, we believe that these transitions are not associated with magnetism of Ni as they are absent in magnetization measurements and also not related to any new phase since we didn't find it from XRD.

To understand the transport mechanisms, we fitted (figure S9) low temperature semiconducting regime using Mott-variable range hopping (Mott-VRH, $\rho \sim \exp(T_M/T)^{0.25}$), and mid-temperature semiconducting regime using Arrhenius ($\rho \sim \exp(E_a/k_BT)$) model for x = 0.115 and 0.155, where $T_M = \frac{18}{\xi^3 N(E_F)k_B}$ is the characteristic temperature for the Mott-VRH with $N(E_F)$ as the electronic density of states (DOS) at the Fermi level, $\xi$ as the localization length, and $E_a$ is the activation energy; obtained fit parameters are listed in table S3. The characteristic temperature $T_M$ increases from x = 0.015 to 0.155. This indicates charge carriers becoming relatively more localized with Ni doping since Mott-VRH model considers the constant DOS or slowly varying function of energy near or at the Fermi level. However, increased localization may be considered rather marginal as the $\rho$ at 5 K increases from ~ 4.4 $\mu\Omega$-m in x = 0 to just ~10.75 $\mu\Omega$-m in x = 0.155 (figure 2a-c). This is reflected from the significantly low value of $T_M$ compared to semiconductor or insulator indicates that the present NSs are highly degenerate or narrow-bandgap semiconductors. This is further supported by obtained low value of $E_a$ ~ 0.44 meV and ~ 1.83 meV for x = 0.115 and 0.155, respectively. The increase in $E_a$ from x = 0.115 to 0.155 suggests that barrier height for charge carriers increases with x in this temperature range. All this could be correlated with the enhanced *ZT* presented later.

Figure 2d shows the $\rho$ of these samples for the T range of 300 K – 580 K. They exhibit three metallic regimes in the *T* range of 300 K to 405 K, 405 K to 425 K, and 425 K to 580 K with a systematic evolution of the slopes and magnitudes in each region. In other words, as *T* increases, $\rho$ of x = 0 increases linearly, while it steeply rises near 410 K to 425 K, and then again increases linearly. An increase in the $\rho$ with *T* shows the metallic behaviour, and the sharp jump near 405 K to 425 K is attributed with structural phase transition from β-Ag$_2$Te to α-Ag$_2$Te[23,26,27,32,33]. The first order nature of phase transition is evident from thermal hysteresis in heating and cooling cycles over a wide temperature range 350 K- 500 K (figure S10). This abrupt increase in $\rho$ across the transition is most likely to be due to a sudden drop in the carrier mobility in line with the earlier report on sharp drop in mobility across the phase transition while *n* continuously increases with increasing temperature[34]. Although $\rho$ for x = 0.015 - 0.155 shows metallic behaviour (in magnitude) similar to x = 0 while its slope changes with x, it rises steeply with *T* in Ag$_{2-x}$Ni$_x$Te for x = 0.055, 0.115 and 0.155 above the transition that broadens and shifts towards higher temperature with increase in x (figure 2d) that is in line with the earlier report[32]. This might be due to decrease in mobility owing to the increase in point defects as x increases. In contrast, $\rho$ for x = 0.205 shows only one MIT (figure 2d, inset), which is attributed to the formation of composite Ag$_2$Te/Ni$_{1.4}$Te. Similar temperature-induced MIT has also been previously reported in Cu@Cu$_2$O nanocomposites[35].

Figure 2e shows negative sign of $S$ for $Ag_{2-x}Ni_xTe$, x = 0 - 0.155 in the $T$ range of 10 K to 300 K. As $T$ decreases, the absolute value of $S$ decreases nearly linearly, approaching zero as $T$ tends to 10 K. This is associated with a slight change in the slope in each data in the mid-temperature regime. This may be correlated with observed broad hump or MITs in $\rho$ (figure 2a-c). Moreover, the absolute value of $S$ increases with x, in line with increase in $\rho$ or decrease in $n$ due to n-type dopant Ni. The negative value of $S$ in the whole $T$ range shows that electrical transport is dominated by electrons associated with Te vacancies that is consistent with EDX. The linear dependence of $S$ on $T$ and its negative value for all the samples suggest that they behave as a small-gap degenerate semiconductor. However, difference in the $S$ values among x = 0 - 0.155 becomes smaller as temperature drops. Notably, the absolute value of $S$ increases linearly with $T$ i.e. $S \propto T$ that indicates their metallic or highly degenerate semiconducting behaviour in this temperature range. This can be explained from the Mott-formula of $S$ for metal or degenerate semiconductor: [1,36],

$$S = -\frac{\pi^2 k_B^2 T}{3e}\left[\frac{d\ln(\sigma(E))}{dE}\right]_{E=E_F} = -\frac{\pi^2 k_B^2 T}{3e}\left[\frac{dn}{ndE} + \frac{1}{\mu}\frac{d\mu}{dE}\right]_{E=E_F}, \qquad (1)$$

where e is the charge of free electron, $E_F$ is the energy at the Fermi level, $n(E)$ is the energy-dependent carrier concentration and $\mu(E)$ is the energy-dependent mobility. It is clear from figure 2e, figure S11 and table S3 that there is systematic change in the slope of $S$ versus $T$ behaviour of x = 0 - 0.155. This is associated with $\left[\frac{d\ln(\sigma(E))}{dE}\right]$ at the Fermi level (eq. 1). The change in the slope of $S$ indicates that energy dependence of the DOS becomes stronger as x increases (table S3). This increase in the DOS at or near the Fermi level directly enhances the absolute $S$. Further, the larger the density of state at the $E_F$, stronger is the energy dependence on $\mu(E)$, enhancing $S$.

The $S$ of a semiconductor in Mott-VRH for 3D can be given as[37]

$$S = \frac{k_B}{2e}\left(\frac{3}{4\pi D(E_F)}\right)^{1/2}\left(\frac{2\alpha}{3}\right)^{3/2}\left(\frac{d\ln D(E)}{dE}\right)(k_B T)^{1/2}, \qquad (2)$$

where $D(E_F)$, $k_B$, and $\alpha$ are DOS at the Fermi level, Boltzmann constant and decay constant. Moreover, the $S \propto T$ behaviour in eq. 1 is in contrast to Mott-VRH conduction found in $\rho$ at low temperature since eq. 2 predicts $S \propto T^{1/2}$ behaviour.

The behaviour of $S$ at high $T$ is completely distinct compared to those at low $T$. Figure 2f shows $S$ over the $T$ range 300 K to 575 K. Its absolute value increases with $T$ with a sudden rise across the structural phase transition (410 K - 425 K) for x = 0. The trends of x = 0.015 - 0.155 are also somewhat similar to that of x = 0 while transition region is broadened relative to it as x increases. They corroborate to those in $\rho$ (figure 2d). Furthermore, $S$ steadily increases at high $T$ for x = 0.115 and 0.155 compared to those of other samples which is consistent with enhancement in $\rho$. As x increases, $S$ increases for x = 0 - 0.155 in the whole $T$ range. The maximum absolute value of $S \sim 194$ μV/K at 580 K in x = 0.155 is closed to the earlier reported value of $\sim$ 190 μV/K at 370 K in Ag$_2$Te nanowires[20].

Note that the value of $S$ increases with increase in Ni doping, which could be possible in two ways. We can see from the equation (2) that the first term in the bracket is associated the carrier density $n$ and second is the carrier mobility $\mu$. Note that the introduction of Ni$^{2+}$ (3d$^8$) nominally to replace Ag$^{1+}$ (4d$^{10}$) in the n-type Ag$_2$Te lattice is likely to increase $n$ due to the extra electrons that act as n-type dopants. The enhancement in $n$ usually reduces the $S$, and hence favour the small value of $S$[4,38]. However, $\mu$ should decrease with increase in Ni doping since it is inversely related to $n$ as $\mu = (ne\rho)^{-1}$ using Drude–Sommerfeld free electron model. We, therefore, believe that the enhancement in $S$ is mainly due to the decrease in $\mu$ or its energy dependence (eq. 1)[4,38]. Furthermore, decrease in $\mu$ corresponds to larger effective mass ($m^*$), and $S \propto \frac{m^*}{n}$. This shows that enhancement in $S$ basically comes from decrease in $\mu$ or increase in $m^*$.

Since the weighted mobility ($\mu_W$) can provide nearly the same information as $\mu$, $\mu_W$ has been used to investigate the charge carrier transport mechanisms as $\mu$[39].

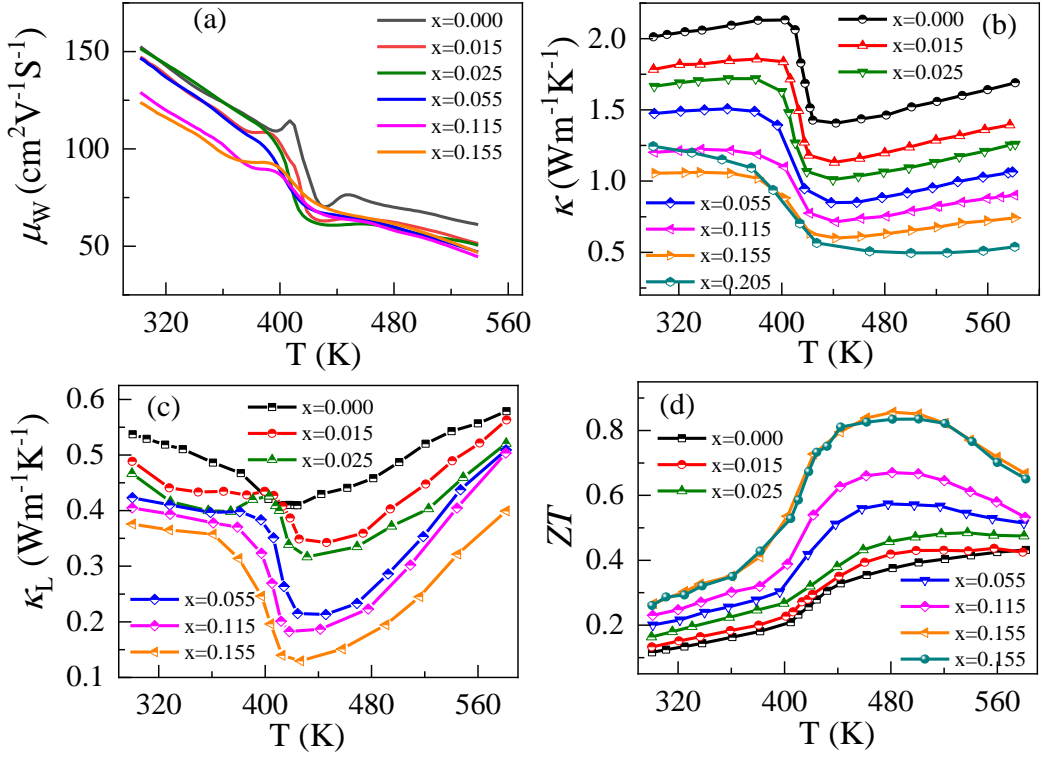

**Figure 3** (a) Calculated weighted mobility ($\mu_W$), (b) total thermal conductivity, $\kappa$, (c) lattice part of the thermal conductivity, $\kappa_L$, and (d) figure of merit (*ZT*) of Ag$_{2-x}$Ni$_x$Te, x = 0 - 0.155; reproducibility checks for x = 0.155 are shown in two curves (purple and dark cyan).

Therefore, we have calculated the $\mu_W$ from $\rho$ and *S* using the equation[39]:

$$\mu_w(cm^2/V.s) = \frac{3h^3\sigma}{8\pi e(2m_e k_B T)^{3/2}} \left[ \frac{exp\left[\frac{|S|}{k_B/e}-2\right]}{1+exp\left[-5\left(\frac{|S|}{k_B/e}-1\right)\right]} + \frac{\frac{3}{\pi^2}\frac{|S|}{k_B/e}}{1+exp\left[5\left(\frac{|S|}{k_B/e}-1\right)\right]} \right] \qquad (3)$$

Figure 3a shows the calculated $\mu_W$ for Ag$_{2-x}$Ni$_x$Te, x = 0 - 0.155. As T increases, $\mu_W$ varies as $\mu_W \sim T^{-0.5}$ below the phase transition and it suddenly drops near the transition, and follows $\mu_W \sim T^{-0.2}$ above the phase transition for all the samples; the sudden drop in $\mu_W$ across the phase transition is consistent with the drop in $\mu$[34]. This behaviour of $\mu_W$ is in contrast to $\mu \sim T^{-1.5}$ reported for Ag$_2$Te[40] that shows the change in the transport behaviour when Ni is doped in the presence of Ag. Notably, $\mu_w$ more or less decreases, specifically at higher dopant concentration with increase in x (figure 3a). The decrease in $\mu_W$ suggests possible enhanced $m^*$, which leads to increase in $S$ with increasing x. The enhancement in $m^*$ with doping of magnetic element (Ni) could be expected as such enhanced $m^*$ and $S$ have previously been reported for Cr-doped $Bi_2Te_3$[38].

Figure 3b depicts $\kappa$ over the $T$ range of 300 K to 580 K for $Ag_{2-x}Ni_xTe$, x = 0 - 0.155. As $T$ increases, $\kappa$ of x = 0 increases linearly above and below the phase transition, like $\rho$, within which (between ~ 400 K-430 K), it drops abruptly. The trends of $\kappa$ for x = 0.015 - 0.155 are somewhat similar to that of x = 0 while transition region becomes broader and shifts with increase in x, which is in line with the earlier reports[23,34]. We have calculated $\kappa_e$ (figure S12a) using Wiedemann-Franz law, $\kappa_e = LT\rho^{-1}$, where $L = 2.44 \times 10^{-8}$ W$\Omega$K$^{-2}$ is the Lorentz number and then $\kappa_L$ is separated from $\kappa$ using $\kappa_L = \kappa - \kappa_e$ (figure 3c). Although $\kappa_e$ shows more or less similar trends as $\kappa$ for x = 0 - 0.025, it decreases with $T$ above the phase transition for x = 0.055 - 0.155 (figure S12a). This decrease in $\kappa_e$ above the phase transition is most likely to be due to the increased electronic scattering in highly doped samples. The overall $\kappa_e$ decreases from x = 0 - 0.155 in the whole $T$ range mainly due to the effective decrease in $n$ with increase in x as it is proportional to $n$. The slope of $\kappa_l$ increases with x that is consistent with decrease in $n$, and hence electron-phonon interaction decreases above the transition. The $\kappa_e$ dominates over $\kappa_l$ in x = 0 - 0.155. Thus, the thermal transport is dominated by electrons, specifically below the transition for x = 0 - 0.155 and above the transition for x = 0 - 0.055 (figure 3b). The $\kappa_L$ mainly decreases with $T$ below the transition for x = 0, 0.055, 0.115 and 0.155, except for x = 0.015 and 0.025 near the phase transition. It starts to increase above the phase transition as $T$ increases suggesting that umklapp scattering or phonon-phonon interactions are not so important in the present case since these interactions usually reduce $\kappa_L$. This enhancement in $\kappa_L$ at high temperature can be correlated with increased electron-phonon interactions due to decrease in $n$ as evident from steep increase in $\rho$, and decrease in $\kappa_e$ above the phase transition for x = 0.055, 0.115 and 0.155.

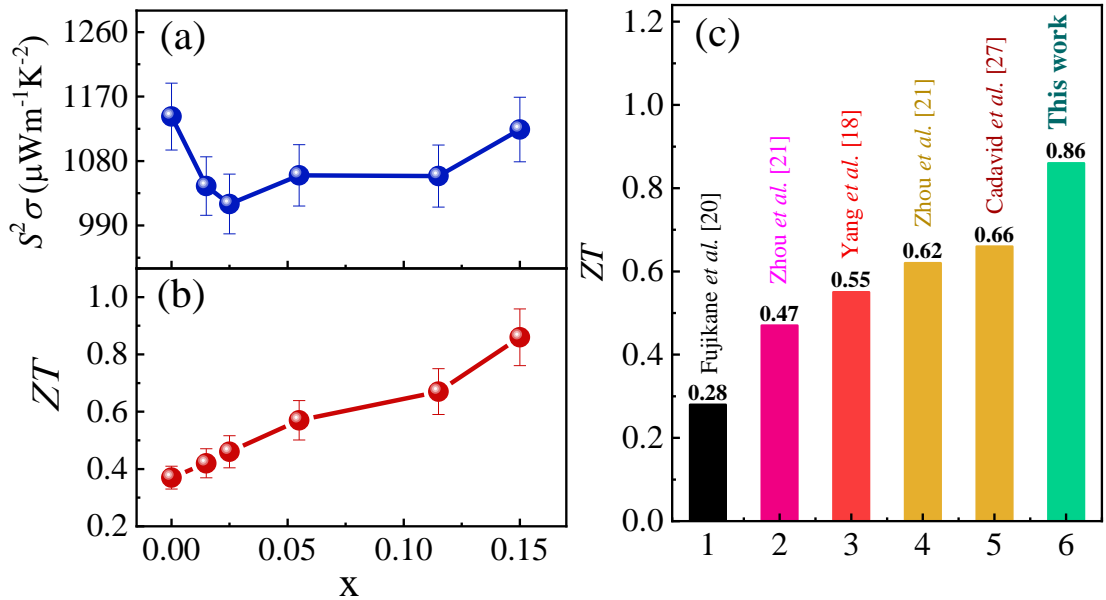

**Figure 4** (a) Power factor and (b) figure of merit (*ZT*) as a function of Ni doping (x), and (c) a comparison of *ZT* for bulk and various $Ag_2Te$ nanostructures.

The substitution of smaller ionic radius $Ni^{2+}$ for the larger ionic radius $Ag^+$ sites increases the lattice distortion as evident from increase in lattice parameter b (figure 1b) and Raman analysis (figure S4). This introduces anharmonicity and affects the covalent bonding or enhance ionic character, in consistent with earlier report[41]. As a result, the overall $\kappa_L$ decreases from x = 0 to 0.155 in $Ag_{2-x}Ni_xTe$. The x = 0.155 shows the lowest value of $\kappa \sim 0.63$ $Wm^{-1}K^{-1}$ and $\kappa_L \sim 0.15$ $Wm^{-1}K^{-1}$ at ~ 481 K which are comparable to the values obtained at 400 K in $Ag_2Te$ nanowires[20]. These values are very low and smaller than those of many well-known potential TE materials[35]. This ultralow value of $\kappa$ could be associated with scattering of heat carriers with point defects and GBs, and anharmonicity. The $Ag_2Te/Ni_{1.4}Te$ nanocomposite x = 0.205 shows $\kappa \sim 0.40$ $Wm^{-1}K^{-1}$ at ~ 481 K that is smaller than x = 0.155, mainly due to additional scattering of heat carriers at the interfaces. The *ZT*s (figure 3d) increase with *T* for $Ag_{2-x}Ni_xTe$, x = 0 - 0.155 with a broad hump each at 480 K and then they start to decrease; the hump in ZT for x = 0 is very broad. Results in x = 0.015 - 0.155 are consistent with earlier report[34]. The value of $ZT$ increases from x = 0 to 0.155 in whole $T$ range except that ZT for x = 0 overtakes marginally that of x = 0.015 at higher $T$. Measurements on the highest $ZT$ valued x = 0.155 were repeated to check its reproducibility that indicates good reproducibility as shown by the almost coinciding two curves (figure 3d); other corresponding parameters have not been shown for clarity. Figure S12b shows the $T$-dependence on PFs; their features are relatively different from $ZT$s.

Figure 4a, b shows PF and $ZT$ versus x, respectively. The PF decreases down to x = 0.025 and increases for higher x; it shows a maximum of ~ 1254 $\mu Wm^{-1}K^{-2}$ at 580 K in x = 0. In contrast, $ZT$ increases with x nearly linearly with its maximum of ~ 0.86 at 482 K in x = 0.155. A comparison of different reported $ZT$s for doped and/or NS $Ag_2Te$ near 480 K is depicted in figure 4c to assess the present result that shows its better value than those of the earlier reports even though our samples were simply compressed at room temperature showing the attractive features of the present report. This elucidates the advantage of combined effect of nanocomposites, nanostructuring, and Ni doping.

**CONCLUSION**

We studied the thermoelectric properties of $Ag_{2-x}Ni_xTe$ (x = 0 - 0.155) nanostructures (NSs) in the temperature range of 10 K to 575 K. The doping-cum-NS evolves their overall metallic behaviour at low temperature to two metal-insulator transitions as x increases. Consequently, the peak figure of merit ($ZT$) of 0.86±0.1 near 480 K for x = 0.155 is achieved that is ~ 83 % enhanced compared to the bulk $Ag_2Te$, and significantly improved over the best value reported for $Ag_2Te$ NS so far at nearly the same temperature. Thus, the simultaneous doping, nanocomposite formation and nanostructuring appears as an effective strategy for improving thermoelectric properties of this material that might well be applicable in similar compounds.

**ASSOCIATED CONTENT**

Synthesis conditions (Table S1); XRD along with Rietveld refinement of $Ag_{2-x}Ni_xTe$, $x = 0 - 0.155$ (Fig. S1); XRD pattern of $x = 0.205$ (Fig. S2); TEM images (Fig. S3), Raman spectra of $x = 0$ and $0.155$ (Fig. S4); XPS of $x = 0$ (Fig. S5); EDX of $x = 0 - 0.155$ (Fig. S6); EDX of $x = 0.055$ at different sites (Fig. S7); elemental analysis for $x = 0 - 0.155$ (Table S2); magnetization data of $x = 0.015$ and $0.025$ (Fig. S8); fitted electrical resistivity using Mott-VRH and Arrhenius model (Fig.S9); fit parameters obtained from Mott-VRH and Arrhenius model (Table S3); electrical resistivity during cooling and heating cycle of $x = 0$ (Fig.S10); fitted Seebeck coefficient using Mott formula (Fig.S11); electronic thermal conductivity and power factor (Fig. S12).

**AUTHOR INFORMATION**

**Corresponding Authors**

**Vikash Sharma**-UGC-DAE Consortium for Scientific Research, University Campus, Khandwa Road, Indore 452001, Madhya Pradesh, India.
http://orcid.org/0000-0002-9277-4411, Email: vikash.sharma@tifr.res.in

**Gunadhor Singh Okram**- UGC-DAE Consortium for Scientific Research, University Campus, Khandwa Road, Indore 452001, Madhya Pradesh, India.
ORCID: http://orcid.org/0000-0002-0060-8556, Email: okram@csr.res.in

**Authors**

**Divya Verma**-Government College Alote, District Ratlam, Madhya Pradesh-457114, India, and Department of chemistry, Vikram University, Ujjain, Madhya Pradesh-456010, India. ORCID:https://orcid.org/0000-0003-4802-6368

**Ranu Bhatt**- Technical Physics Division, Bhabha Atomic Research Centre, Mumbai-400085, India.

**Pankaj K. Patro**- Powder Metallurgy Division, Bhabha Atomic Research Centre, Vashi Complex, Mumbai-400703, India.

**Present Addresses**

†Department of Condensed Matter Physics & Materials Science, Tata Institute of Fundamental Research, Homi Bhabha Road, Mumbai-400005, India.

## ACKNOWLEDGEMENTS

Authors are thankful to Dr. R. J. Choudhary, Dr. Vasant Sathe, Dr. Alok Banerjee and Mr. Vinay Ahire, UGC-DAE Consortium for Scientific Research, Indore, India for providing XRD, XPS, Raman, magnetization and EDX facilities and CIL, Harisingh Gour University, Sagar for providing TEM and FESEM data. Vikash Sharma gratefully acknowledge Er. Vipin Kumar, Council of Scientific & Industrial Research (CSIR)-Central Electronics Engineering Research Institute (CEERI), Pilani, Rajasthan, India, Dr. Gaurav Sharma, IISER Mohali, India and Dr. Gyanendra Panchal, Helmholtz Zentrum Berlin, Germany for help and support.

Improvement of Thermoelectric Performance in Sb2Te3/Te Composites. *Phys. Rev. Mater.* **2022**, *6*, 035401.

(41) Vaney, J.; Yamini, S. A.; Takaki, H.; Kobayashi, K.; Kobayashi, N.; Mori, T. Magnetism-Mediated Thermoelectric Performance of the Cr-Doped Bismuth Telluride Tetradymite. *Mater. Today Phys.* **2019**, *9*, 100090.